\documentclass[pra,twocolumn,superscriptaddress,showpacs]{revtex4}
\usepackage{graphicx}
\usepackage{amsmath}
\usepackage{amssymb,bm}
\usepackage{bm,hyperref}
\usepackage{CJK}

\begin{document}
\begin{CJK*}{UTF8}{gbsn}
\title{Superfluidity of pure spin current in ultracold Bose gases}
\author{Qizhong Zhu (朱起忠)}
\affiliation{International Center for Quantum Materials, School of Physics, Peking University, Beijing 100871, China}
\author{Qing-feng Sun (孙庆丰)}
\affiliation{International Center for Quantum Materials, School of Physics, Peking University, Beijing 100871, China}
\affiliation{Collaborative Innovation Center of Quantum Matter, Beijing 100871, China}
\author{Biao Wu (吴飙)}
\affiliation{International Center for Quantum Materials, School of Physics,
Peking University, Beijing 100871, China}
\affiliation{Collaborative Innovation Center of Quantum Matter, Beijing 100871, China}
\affiliation{Wilczek Quantum Center, College of Science, Zhejiang University of Technology,  Hangzhou 310014, China}
\date{\today}

\begin{abstract}
We study the superfluidity of a pure spin current that is a spin current without
mass current. We examine two types of pure spin currents, planar and circular,
in spin-1 Bose gas. For the planar current, it is usually unstable,
but  can be  stabilized by the quadratic Zeeman effect.  The circular current can be
generated with spin-orbit coupling.  When the spin-orbit coupling
strength is weak,  we find that the circular pure spin current is
the ground state of the system  and thus a super-flow.
We discuss the experimental schemes to realize and detect  a pure spin current.
\end{abstract}
%05.30.Jp	   Boson systems
%03.75.Mn  Multicomponent condensates; spinor condensates
%03.75.Kr   Dynamic properties of condensates; collective and hydrodynamic excitations, superfluid flow
%71.70.Ej	   Spin-orbit coupling, Zeeman and Stark splitting, Jahn-Teller effect
\maketitle
\end{CJK*}

\section{Introduction}
Since the experimental realization of Bose-Einstein condensation in optical traps,
much effort has been devoted to the study of spinor
superfluid \cite{hall1,hall2,matthews,stenger,ho,ohmi,review}. With the
degree of freedom of spin, spinor superfluid has much richer phases than scalar
superfluid as it has both superfluid order and spin textures.
However, most spinor superfluids studied till now carry both spin current and
mass current. It will be interesting to see whether these two currents can
decouple and further whether a  pure spin current which has no mass current
can flow frictionlessly.  Our motivation also originates
from condensed matter physics, in which the concept of spin
superconductor \cite{sun2,sun3},
formed by the Cooper-like pairs of electrons and holes and carrying spin
super-current, is proposed. It is interesting to have these ideas realized
in the field of cold atoms.

In this work we focus on the unpolarized spin-1 Bose-Einstein condensate (BEC), where the pure spin current
can be generated by applying a small magnetic gradient.  It was found
in Ref. \cite{fujimoto} that  a planar pure spin current  in such
a system is always unstable as the $m=1,-1$ components can collide
into the $m=0$ component destroying the spin current. We find that
the pure spin current can be stabilized with the quadratic Zeeman effect
and become a super-flow.  Furthermore, we propose a scheme to create pure
spin current at the ground state, thus free from the issue of instability. Our scheme
utilizes the spin-orbit coupling. Specifically, we study a spin-1 BEC
with Rashba spin-orbit coupling confined in a two-dimensional harmonic trap,
and numerically find the ground state of the system. For antiferromagnetic
interactions, opposite vortices appear in the $m=1,-1$ components with equal
amplitude when the spin-orbit coupling is weak. Such a state carries
pure spin current and no mass current. This spin current is a super-flow as it is
the ground state and must be stable.

We note that there has been a lot of theoretical and experimental work on the counterflow of two species BEC \cite{law,kuklov,yukalov,takeuchi,engels,hamner,fujimoto,ishino,jltp,abad,kurn,cherng}.
For two miscible BECs with counterflow, it is found that there is a critical relative speed between the two species, beyond which the state is dynamically unstable \cite{law,kuklov,yukalov,takeuchi,engels,hamner,ishino,jltp,abad}. It is shown that
the instability can lead to proliferation of solitons \cite{engels,hamner} and
quantum turbulence \cite{takeuchi}.  This kind of counterflow is very similar
to a spin current but it is not for two reasons: (1) Theoretically, if we regard
the two species as two components of a pseudo-spin, this  pseudo-spin has no SU(2) rotational
symmetry as the number of bosons in each species is conserved. (2) Experimentally,
it is hard to control the number of bosons in each component to create a spin current
that has no mass current.

The paper is organized as follows. In Sec. \ref{planar}, we first study the stability
of a spin-1 planar counterflow. We identify the mechanisms associated with
the instabilities, and find that the quadratic Zeeman effect  can stabilize such a planar
flow. We then study the similar situation in the circular geometry
in Sec. \ref{circular}. The pure spin current consists of a vortex and
anti-vortex in the $m=-1,1$ components. The experimental schemes
to realize the stable pure spin current is discussed in Sec. \ref{expe}.
Finally, we briefly summarize our main results in Sec. \ref{conclusion}.

\section{Planar flow}\label{planar}
The dynamics of a spin-1 BEC in free space is governed by the
mean field Gross-Pitaevskii (GP) equation \cite{review},
\begin{equation}
\label{ }
i\hbar\frac{\partial}{\partial t}\psi_m=-\frac{\hbar^2\nabla^2}{2M}\psi_m+c_0 \rho\psi_m+c_2\sum_{n=-1}
^1\mathbf{s}\cdot\mathbf{S}_{mn}\psi_n,
\end{equation}
where $\psi_m$ ($m=1,0,-1$) are the components of the macroscopic wave function.
$\rho=\sum_{m=-1}^1|\psi_m|^2$ is the total density, ${\bf s}_i=\sum_{mn}\psi_m^*(S_i)_{mn}\psi_n$
is the spin density vector and ${\bf S}=(S_x, S_y, S_z)$ is the spin operator vector with $S_i$ ($i=x,y,z$)
being the three Pauli matrices in the spin-1 representation.
The collisional interactions include a spin-independent
part $c_{0}=4\pi\hbar^{2}(a_{0}+2a_{2})/3M$ and a spin-dependent
part $c_{2}=4\pi\hbar^{2}(a_{2}-a_{0})/3M$, with $a_{f}(f=0,2)$
being the $s$-wave scattering length for spin-1 atoms in the
symmetric channel of total spin $f$.

We consider a spin current state of the above GP equation with the form
\begin{equation}
\label{scurrent}
\psi=\sqrt{\frac{n}{2}}\left(
\begin{array}{c}
e^{i\mathbf{k}_1\cdot{\bf r}}\\  0\\
e^{i\mathbf{k}_2\cdot{\bf r}}
\end{array}
\right)\,,
\end{equation}
where $n$ is the density of the uniform BEC.
The requirement of equal chemical potential leads to $|\mathbf{k}_1|=|\mathbf{k}_2|$.
In the case where $\mathbf{k}_1=-\mathbf{k}_2$,  this state carries a pure spin current:
the total mass current is zero as  it has equal mass counterflow while the spin current is nonzero.

It is instructive to first consider the special case when there is no counterflow, i.e., ${\bf k}_1={\bf k}_2=0$.
The excitation spectra are found to be $\epsilon^0=\sqrt{2c_2n\epsilon_q+\epsilon_q^2}$ and $\epsilon_1^{\pm1}=\sqrt{2c_0n\epsilon_q+\epsilon_q^2}$, $\epsilon_2^{\pm1}=\sqrt{2c_2n\epsilon_q+\epsilon_q^2}$, respectively,
with $\epsilon_q=\hbar^2q^2/2M$.
So for antiferromagnetic interaction ($c_0>0, c_2>0$), all branches of the spectra are real
and  there is a double degeneracy in one branch of the spectra. The phonon excitations give two sound velocities, $\sqrt{n c_i/M}$ ($i=0,2$), corresponding to the speeds of density wave and spin wave, respectively. However, the existence of phonon excitation
does not mean that  the pure spin current (${\bf k}_1={\bf k}_2\neq 0$) is a super-flow
as we can not obtain the current with ${\bf k}_1={\bf k}_2\neq 0$ from the state
with ${\bf k}_1={\bf k}_2= 0$ by a Galilean transformation.

The stability of the spin current  has been studied
in Ref. \cite{fujimoto} for the  case $\mathbf{k}_1=-\mathbf{k}_2\neq 0$.
It is found that, for  the antiferromagnetic interaction case ($c_0>0, c_2>0$), the excitation
spectrum of the $m=0$ component always has nonzero imaginary part
in the long wavelength limit as long as there is counterflow between the two components, and the imaginary excitations in the $m=1,-1$ components only appear for a large enough relative velocity $v_1=2\sqrt{n c_2/M}$. For the ferromagnetic
interaction case ($c_0>0, c_2<0$), both excitation spectra of the $m=0$ and $m=1,-1$ components have nonzero imaginary parts for any relative velocity. This means that the pure spin current cannot be stable in any cases.

For the general non-collinear case (${\bf k}=\frac{{\bf k}_1+{\bf k}_2}{2}\neq0$) and antiferromagnetic interaction, the excitation spectrum for the $m=0$ component is found to be
\begin{equation}
\label{ }
\epsilon^0=\sqrt{\left(\epsilon_q+\frac{\hbar^2}{2M}\left(|\mathbf{k}|^2-|\mathbf{k}_1|^2\right)+c_2n\right)^2-c_2^2n^2}.
\end{equation}
We see here that as long as the momenta of the two components are not exactly parallel, i.e., $\mathbf{k}_1$ is not exactly equal to $\mathbf{k}_2$, then $|\mathbf{k}|<|\mathbf{k}_1|$, and
there is always dynamical instability for the long wavelength excitations.

Therefore, the spin current in Eq. (\ref{scurrent}) is generally unstable and not a super-flow.
This instability  originates from the interaction process described by
$\psi_0^{\dagger}\psi_0^{\dagger}\psi_1\psi_{-1}$ in the second
quantized Hamiltonian. This energetically favored process converts two
particles in the $m=1,-1$ components, respectively,  into two stationary particles
in the $m=0$ component. To suppress such a process and achieve a stable
pure spin current, one can utilize the quadratic Zeeman effect.
With the quadratic Zeeman effect of negative coefficient, the Hamiltonian
adopts an additional term $\lambda m^2$ ($\lambda<0$ and $m=1,0,-1$).
This term does not change the energy of the $m=0$
component, but lowers the energy of the other two components $m=1,-1$.
As a result, there arises a barrier for two atoms in the $m=1,-1$ components scattering to the $m=0$ component, and the scattering process  is thus suppressed.

The above intuitive argument can be made more rigorous and quantitative.
Consider the case $\mathbf{k}_1=-\mathbf{k}_2$. With the quadratic Zeeman term,
the  excitation spectrum for the $m=0$ component changes to
\begin{equation}
\label{ }
\epsilon^0=\sqrt{\left(\epsilon_q-\frac{\hbar^2 |\mathbf{k}_1|^2}{2M}+c_2n-\lambda\right)^2-c_2^2n^2}.
\end{equation}
So as long as $-\lambda-\hbar^2 |\mathbf{k}_1|^2/2M>0$, long wavelength excitations will be stable for the $m=0$ component. From the excitation
of the $m=0$ component, one can obtain a critical relative velocity of the spin current,
$v_0=2\sqrt{-2\lambda/M}$.  There is  another nonzero critical velocity $v_1=2\sqrt{n c_2/M}$ determined by the
excitations of the $m=1,-1$ components.  The overall critical velocity of the system is the smaller
one of $v_0$ and $v_1$.  Therefore, below the critical relative velocity $v_{\rm c}={\rm min}\{v_0,v_1\}$, the pure spin current is stable and a super-flow. The experimental scheme
to realize such a Zeeman effect will be discussed  in Sec. \ref{expe}.

\section{Circular flow}\label{circular}
In the cylindrical geometry, we consider a pure spin current formed by two vortices
 with opposite circulation
in the $m=1,-1$ components. From similar arguments, one can expect that interaction will make
such a current unstable.  Inspired by the quadratic Zeeman effect method above,
we propose to use spin-orbit coupling to stabilize it. The spin-orbit coupling can be
viewed as a momentum-dependent effective magnetic field that exerts only on
the $m=1,-1$ components. Therefore, it is possible that spin-orbit coupling lowers
the energy of $m=1,-1$ components, and consequently suppresses the interaction process
leading to the instability.

The model of spin-1 BEC subject to Rashba spin-orbit coupling
can be described by the following energy functional,
\begin{align}
\label{energy}
\mathcal{E}\left[\psi_{\alpha}\right]= & \int d\mathbf{r}\Bigg\{\sum_{\alpha}\frac{\hbar^{2}|\nabla\psi_{\alpha}|^{2}}{2M}+\rho V(r)+\frac{c_{0}}{2}\rho^{2}
+\frac{c_{2}}{2}\mathbf{s}^{2} \nonumber \\
& +\gamma\langle S_xp_y-S_yp_x\rangle\Bigg\},
\end{align}
where $\rho$ is the density, $V(r)=\frac{1}{2}M\omega^2(x^2+y^2)$ is the trapping potential, and $\gamma$ is the strength of spin-orbit coupling. $\langle\cdots\rangle$ is the expectation value taken with respect to the three component wave function
$\psi=(\psi_1,\psi_0,\psi_{-1})^T$. The strength of the spin-orbit coupling $\gamma$ defines a characteristic length
$a_{\rm soc}=\hbar/M\gamma$, and can be rescaled to be dimensionless with respect to the harmonic oscillator length
$a_{\rm h}=\sqrt{\hbar/M\omega}$. Then we characterize the strength of spin-orbit coupling with the dimensionless
quantity $\kappa=a_{\rm h}/a_{\rm soc}=\gamma\sqrt{M/\hbar\omega}$.
The spin-orbit coupling of Rashba type here can be generated in various ways, which will be discussed in the next section.

The above model can describe a spin-1 BEC of $^{23}$Na confined in a
two-dimensional harmonic trap. Assume the atom number is about $10^6$.
Using the estimate of scattering lengths $a_0=50a_B$, $a_2=55a_B$ \cite{epjd},
with $a_B$ being the Bohr radius, the ground state of spin-1 $^{23}$Na should be
antiferromagnetic because $c_0>0, c_2>0$ \cite{ho}.
Previous studies of spin-1 BEC with Rashba spin-orbit coupling mostly focus on
the strong spin-orbit coupling regime, where the ground
state is found to be the plane wave phase or the stripe phase, for ferromagnetic interaction and antiferromagnetic interaction, respectively \cite{zhai}.
Here we are interested in the antiferromagnetic interaction case and the
weak spin-orbit coupling regime ($\kappa\ll 1$), and calculate the ground
state wave function of the energy functional with the method of imaginary
time evolution.

\begin{figure}[t]
\begin{center}
\includegraphics[width=8.5cm]{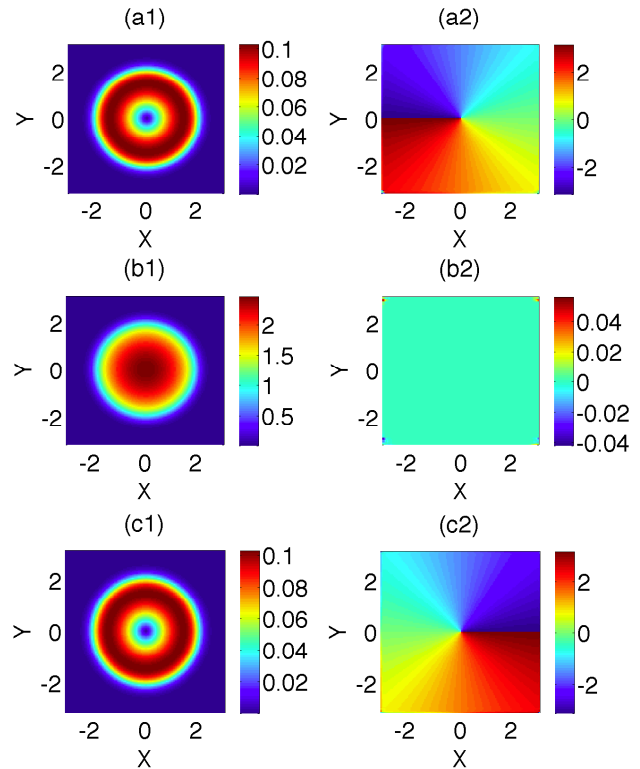}
\caption{(color online) Amplitudes (a1,b1,c1) and phase angles (a2,b2,c2) of the three component wave function
$\psi=(\psi_1,\psi_0,\psi_{-1})^T$ at the ground state of Hamiltonian (\ref{energy}) for a BEC of $^{23}$Na confined
in a 2D harmonic trap. The particle number is $10^6$, the frequency of the trap is $2\pi\times42$ Hz, and the dimensionless spin-orbit coupling strength is $\kappa=0.04$. The units of the $X$ and $Y$ axes are $a_{\rm h}$.}
\label{wf}
\end{center}
\end{figure}

We find that when the spin-orbit coupling is weak ($\kappa\ll 1$), the ground state wave function
has the form
\begin{equation}\label{ground}
\psi=\left(\begin{array}{c}
\chi_{1}(r)e^{-i\phi}\\
\chi_{0}(r)\\
\chi_{-1}(r)e^{i\phi}
\end{array}\right),
\end{equation}
with $\chi_{1}(r)=-\chi_{-1}(r)$ and all $\chi_i$ real. The ground state is
shown in Fig. \ref{wf}. Such a ground state consists of an anti-vortex
in the $m=1$ component and a vortex in the $m=-1$ component.
The $m=0$ component does not carry angular momentum. Since $|\psi_1|=|\psi_{-1}|$,
the net mass current vanishes.

The wave function in Eq. (\ref{ground}) can be understood in the single particle level.
In terms of the ladder operators of spin and angular momentum, the spin-orbit coupling term reads
\begin{equation}
\mathcal{H}_{\rm soc}=\frac{\gamma\sqrt{M\hbar\omega}}{2}\left[S_+\left(\hat{a}_R-\hat{a}_L^{\dagger}
\right)+S_-\left(\hat{a}_R^{\dagger}-\hat{a}_L\right)\right],
\end{equation}
where $S_{\pm}$ is the ladder operator of spin, and $\hat{a}_{L(R)}^{\dagger}$ is the creation operator of the
left (right) circular quanta \cite{cohen}. When the spin-orbit coupling is very weak ($\kappa\ll 1$), its effect can be accounted
for in a perturbative way. From the ground state $\Psi^{(0)}=|0,0\rangle$, the first order correction to the
wave function for small $\gamma$ is given by
\begin{align}
\Psi^{(1)}=&\frac{\gamma\sqrt{M\hbar\omega}}{2\hbar\omega}\left(-S_+\hat{a}_L^{\dagger}
+S_-\hat{a}_R^{\dagger}\right)|0,0\rangle \nonumber \\
=&\frac{\kappa}{2}\left(-|1,-1\rangle+|-1,1\rangle\right),
\end{align}
where $|m_s,m_o\rangle$ denotes a state with spin quantum number $m_s$ and orbital magnetic quantum number
$m_o$. One immediately sees that $\psi_1$ has angular momentum $-\hbar$ and $\psi_{-1}$ has angular momentum $\hbar$.
Besides, the amplitudes of both $\psi_1$ and $\psi_{-1}$ are proportional to $\kappa$.

There exits a continuity equation for spin density and spin current, which  is
\begin{equation}
\frac{d}{d t}\left(\psi^{\dagger}{\bf S}_\mu\psi\right)+\nabla\cdot\mathbf{J}_{\mu}^{s}=0.
\end{equation}
The spin current density tensor $\mathbf{J}_{\mu}^{s}$ ($\mu=x, y, z$ denotes the spin component) is defined as \cite{aref1,aref2}
\begin{align}
\label{def}
\mathbf{J}_{\mu}^{s}=&\frac{1}{2}\left\{\psi^{\dagger}S_\mu{\bf v}\psi+{\rm c.c.}\right\} \nonumber \\
 =&\frac{1}{2}\left\{\sum_{m,n,l}\psi_m^*\left(S_\mu\right)_{mn}{\bf v}_{nl}\psi_l+{\rm c.c.}\right\},
\end{align}
where
\begin{equation}
{\bf v}_{nl}=\frac{{\bf p}}{M}+\gamma\left(\hat{z}\times{\bf S}_{nl}\right),
\end{equation}
and c.c. means the complex conjugate. The second part in ${\bf v}_{nl}$  is
induced by the spin-orbit coupling.

\begin{figure}
\begin{center}
\includegraphics[width=8.5cm]{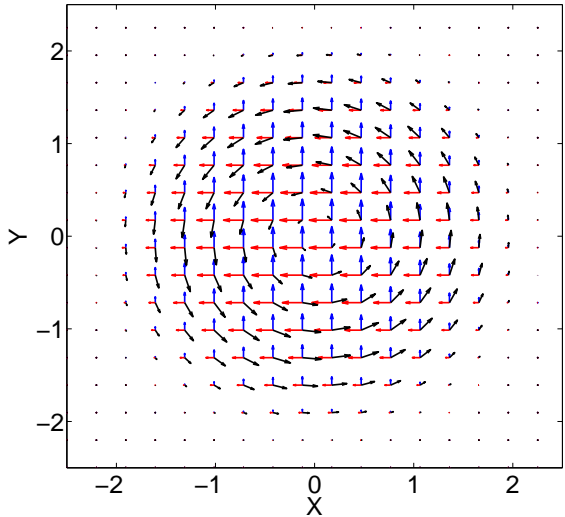}
\caption{(color online) Distribution of the spin current densities $\mathbf{J}_{x}^{s}$ (blue arrow), $\mathbf{J}_{y}^{s}$ (red arrow) and
$\mathbf{J}_z^s$ (black arrow)
of the ground state shown in Fig. \ref{wf}. The length of the arrows represents the
strength of the spin current. The arrow length of different colors is not to scale. $\kappa=0.04$.
The units of the $X$ and $Y$ axes are $a_{\rm h}$.}
\label{current}
\end{center}
\end{figure}

By the definition in Eq. (\ref{def}), the spin current density carried by
the ground state (\ref{ground}) is
\begin{eqnarray}
\label{spincurrent}
\mathbf{J}_{x}^{s}=&\gamma\sin2\phi|\psi_1|^2\hat{x}+\gamma\left(|\psi_0|^2+2|\psi_1|^2\sin^2\phi\right)\hat{y}, \nonumber\\
\mathbf{J}_{y}^{s}=&-\gamma\left(|\psi_0|^2+2|\psi_1|^2\cos^2\phi\right)\hat{x}-\gamma\sin2\phi|\psi_1|^2\hat{y}, \nonumber\\
\mathbf{J}_{z}^{s}
=&\left(-\frac{2\hbar|\psi_{1}|^{2}}{Mr}+\sqrt{2}\gamma|\psi_1\psi_0|\right)\hat{\phi}.
\end{eqnarray}
From both analytical and numerical results of the wave function, $|\psi_1|\ll|\psi_0|$, so $\mathbf{J}_{x}^{s}$
roughly points in the $y$ direction, while $\mathbf{J}_{y}^{s}$ almost points in the $-x$ direction.
$\mathbf{J}_{z}^{s}$
represents a flow whose amplitude has rotational symmetry.
 From the numerical results shown in
Fig. \ref{current}, we see that $\mathbf{J}_{z}^{s}$ is a counter-clockwise flow.
The amplitudes of $\mathbf{J}_{x}^{s}$ and $\mathbf{J}_{y}^{s}$ are of the same order, both proportional to $\kappa$,
while that of $\mathbf{J}_{z}^{s}$, proportional to $\kappa^2$, is much smaller.
It is evident that the state in Eq. (\ref{ground}) carries no mass current and only pure spin current. Since the spin current is in the ground state, it
must be stable.  In this way,  we have realized a superfluid of pure spin current,
or a pure spin super-current.

\section{experimental schemes}\label{expe}
In this section, we propose the experimental schemes to generate and detect
the pure spin currents discussed in Sec. \ref{planar} and Sec. \ref{circular}.

The planar pure spin current can be easily generated. By  applying a magnetic field gradient,
the two components $m=1,-1$  will be accelerated in opposite directions and a pure
spin current is generated as done in Refs. \cite{engels,hamner}.  To stabilize this
spin current, one needs to generate the quadratic  Zeeman effect. We apply an  oscillating magnetic field $B\sin\omega t$ with the frequency $\omega$ being much larger than
the characteristic frequency of the condensate, e.g., the chemical potential $\mu$.
The time averaging removes the linear Zeeman effect; only the quadratic Zeeman effect
remains.  The coefficient of the quadratic  Zeeman effect from the second-order
perturbation theory is given by $\lambda=\left(g\mu_{\rm B} B\right)^2/\Delta E_{\rm hf}$, where $g$ is the Land\'e $g$-factor of the atom, $\mu_{\rm B}$ is the Bohr
magneton, and $\Delta E_{\rm hf}$ is the hyperfine energy splitting \cite{ueda}. For the $F=2$ manifold of $^{87}$Rb, $\Delta E_{\rm hf}<0$, so the coefficient of the quadratic
Zeeman effect is negative.

The circular flow  in Sec. \ref{circular} may find prospective realizations in
two different systems: cold atoms and exciton BEC. In cold atoms, we consider a
system consisting of a BEC of $^{23}$Na confined in a pancake trap, where the
confinement in the $z$ direction is so tight  that one can treat the system effectively
as two dimensional. The spin-orbit coupling can be induced by two different methods.
One is by the exertion of a strong external electric field ${\bf E}$ in the $z$ direction. Due to the relativistic effect,
the magnetic moment of the atom will experience a weak spin-orbit coupling,
where the strength $\gamma=g\mu_B|{\bf E}|/Mc^2$. Here $M$ is the atomic mass
and $c$ is the speed of light.
For weak spin-orbit coupling (small $\gamma$), the fraction of atoms in the $m=1,-1$ components
is proportional to $\gamma^2$.
For an experimentally observable fraction of atoms, e.g., $0.1\%$ of $10^6$ atoms,
using the typical parameters of $^{23}$Na BEC,
the estimated electric field is of the same order of magnitude as the vacuum breakdown field.
For atoms with smaller mass or larger magnetic moment, the required electric field can be lowered.
Another method of realizing spin-orbit coupling
is to exploit the atom laser interaction, where strong spin-orbit coupling can be
created in principle \cite{socreview}.
In exciton BEC systems, as the effective mass of exciton is much smaller than that of atom, the required
electric field is four to five orders of magnitude smaller, which is quite feasible in experiments \cite{exciton1,exciton2,exciton3,exciton4}.

The vortex and anti-vortex in the $m=1,-1$ components can be detected by the method
of time of flight.
First one can split the three spin components with the Stern-Gerlach effect.
The appearance of vortex or anti-vortex in the $m=1,-1$ components is
signaled by a ring structure in the time of flight image. After a sufficiently long
time of expansion, the ring structure should be clearly visible.

\section{conclusion}\label{conclusion}
In summary, we have studied the stability of a pure spin current of a spin-1 BEC.
In the planar flow, the system always suffers from dynamical instability. The origin of the
instability is the interaction process that
converts two  particles in the $m=1,-1$ components into the $m=0$ component. Based on this,
we propose a method to stabilize the pure spin current by utilizing the quadratic  Zeeman effect.
In the circular flow, we have proposed to use spin-orbit coupling to make the pure spin current stable.
For weak spin-orbit coupling,  we have found that the ground state of the system is a
superfluid of pure spin current. The experimental schemes to
realize and detect these pure spin currents have been discussed.

\section*{Acknowledgements} This work is supported by the NBRP of China (2013CB921903,2012CB921300) and the NSF of China (11274024,11274364,11334001,11429402).

\end{document}